\pdfoutput=1
\documentclass[12pt,a4paper]{article}
\usepackage{warpcol}
\usepackage{booktabs}
\usepackage{ifthen} 
\newboolean{pdflatex}
\setboolean{pdflatex}{true} 

\newboolean{articletitles}
\setboolean{articletitles}{true} 

\newboolean{uprightparticles}
\setboolean{uprightparticles}{false} 

\def\paperauthors{LHCb collaboration} 
\def\paperasciititle{First observation of the decay Lb -> etac(1S) p K-} 
\def\papertitle{First observation of the decay \\ $\decay{\Lb}{\etac(1S)\proton\Km}$} 
\def\paperkeywords{{High Energy Physics}, {LHCb}} 
\def\papercopyright{\the\year\ CERN for the benefit of the LHCb collaboration} 
\def\paperlicence{CC BY 4.0 licence}
\def\paperlicenceurl{https://creativecommons.org/licenses/by/4.0/}

\usepackage[top=1in, bottom=1.25in, left=1in, right=1in]{geometry}

\usepackage{microtype}
\usepackage{lineno}  
\usepackage{xspace} 
\usepackage{caption} 

\usepackage{graphicx}  
\usepackage{color}
\usepackage{colortbl}
\graphicspath{{./figs/}} 

\usepackage{amsmath} 
\usepackage{amssymb}
\usepackage{amsfonts}
\usepackage{upgreek} 

\newcommand*\patchAmsMathEnvironmentForLineno[1]{%
\expandafter\let\csname old#1\expandafter\endcsname\csname #1\endcsname
\expandafter\let\csname oldend#1\expandafter\endcsname\csname
end#1\endcsname
 \renewenvironment{#1}%
   {\linenomath\csname old#1\endcsname}%
   {\csname oldend#1\endcsname\endlinenomath}%
}
\newcommand*\patchBothAmsMathEnvironmentsForLineno[1]{%
  \patchAmsMathEnvironmentForLineno{#1}%
  \patchAmsMathEnvironmentForLineno{#1*}%
}
\AtBeginDocument{%
\patchBothAmsMathEnvironmentsForLineno{equation}%
\patchBothAmsMathEnvironmentsForLineno{align}%
\patchBothAmsMathEnvironmentsForLineno{flalign}%
\patchBothAmsMathEnvironmentsForLineno{alignat}%
\patchBothAmsMathEnvironmentsForLineno{gather}%
\patchBothAmsMathEnvironmentsForLineno{multline}%
\patchBothAmsMathEnvironmentsForLineno{eqnarray}%
}

\usepackage{hyperxmp}

\usepackage[pdftex,
            pdfauthor={\paperauthors},
            pdftitle={\paperasciititle},
            pdfkeywords={\paperkeywords},
            pdfcopyright={Copyright (C) \papercopyright},
            pdflicenseurl={\paperlicenceurl}]{hyperref}

\usepackage[colorinlistoftodos,textsize=scriptsize]{todonotes}

\usepackage[bottom,flushmargin,hang,multiple]{footmisc}

\usepackage[all]{hypcap} 

\usepackage{xspace} 
\usepackage{upgreek}

\def\MagUp {\mbox{\em Mag\kern -0.05em Up}\xspace}

\ifthenelse{\boolean{uprightparticles}}%
{

 \def\Peta        {\ensuremath{\upeta}\xspace}

 \def\Pmu         {\ensuremath{\upmu}\xspace}

 \def\Ppi         {\ensuremath{\uppi}\xspace}

 \def\Pphi        {\ensuremath{\upphi}\xspace}

 \def\Ppsi        {\ensuremath{\uppsi}\xspace}

 \def\PDelta      {\ensuremath{\Delta}\xspace}                 
 \def\PXi         {\ensuremath{\Xi}\xspace}                 
 \def\PLambda     {\ensuremath{\Lambda}\xspace}                 
 \def\PSigma      {\ensuremath{\Sigma}\xspace}                 
 \def\POmega      {\ensuremath{\Omega}\xspace}                 
 \def\PUpsilon    {\ensuremath{\Upsilon}\xspace}

 \def\PB      {\ensuremath{\mathrm{B}}\xspace}                 
                  
 \def\PD      {\ensuremath{\mathrm{D}}\xspace}

 \def\PJ      {\ensuremath{\mathrm{J}}\xspace}                 
 \def\PK      {\ensuremath{\mathrm{K}}\xspace}

 \def\Pb      {\ensuremath{\mathrm{b}}\xspace}                 
 \def\Pc      {\ensuremath{\mathrm{c}}\xspace}                 
 \def\Pd      {\ensuremath{\mathrm{d}}\xspace}

 \def\Pi      {\ensuremath{\mathrm{i}}\xspace}

 \def\Pp      {\ensuremath{\mathrm{p}}\xspace}

 \def\Ps      {\ensuremath{\mathrm{s}}\xspace}                 
                  
 \def\Pu      {\ensuremath{\mathrm{u}}\xspace}

 \def\thebaroffset{0.0em}
}
{

 \def\Peta        {\ensuremath{\eta}\xspace}

 \def\Pmu         {\ensuremath{\mu}\xspace}

 \def\Ppi         {\ensuremath{\pi}\xspace}

 \def\Pphi        {\ensuremath{\phi}\xspace}

 \def\Ppsi        {\ensuremath{\psi}\xspace}                 
                  
 \mathchardef\PDelta="7101
 \mathchardef\PXi="7104
 \mathchardef\PLambda="7103
 \mathchardef\PSigma="7106
 \mathchardef\POmega="710A
 \mathchardef\PUpsilon="7107
                  
 \def\PB      {\ensuremath{B}\xspace}                 
                  
 \def\PD      {\ensuremath{D}\xspace}

 \def\PJ      {\ensuremath{J}\xspace}                 
 \def\PK      {\ensuremath{K}\xspace}

 \def\Pb      {\ensuremath{b}\xspace}                 
 \def\Pc      {\ensuremath{c}\xspace}                 
 \def\Pd      {\ensuremath{d}\xspace}

 \def\Pi      {\ensuremath{i}\xspace}

 \def\Pp      {\ensuremath{p}\xspace}

 \def\Ps      {\ensuremath{s}\xspace}                 
                  
 \def\Pu      {\ensuremath{u}\xspace}

 \def\thebaroffset{0.18em}
}
\newcommand{\offsetoverline}[2][\thebaroffset]{\kern #1\overline{\kern -#1 #2}}%

\makeatletter
\ifcase \@ptsize \relax
  \newcommand{\miniscule}{\@setfontsize\miniscule{4}{5}}
\or
  \newcommand{\miniscule}{\@setfontsize\miniscule{5}{6}}
\or
  \newcommand{\miniscule}{\@setfontsize\miniscule{5}{6}}
\fi
\makeatother

\DeclareRobustCommand{\optbar}[1]{\shortstack{{\miniscule (\rule[.5ex]{1.25em}{.18mm})}
  \\ [-.7ex] $#1$}}

\def\mup        {{\ensuremath{\Pmu^+}}\xspace}
\def\mun        {{\ensuremath{\Pmu^-}}\xspace} 

\def\uquark    {{\ensuremath{\Pu}}\xspace}

\def\dquark    {{\ensuremath{\Pd}}\xspace}

\def\squark    {{\ensuremath{\Ps}}\xspace}

\def\cquark    {{\ensuremath{\Pc}}\xspace}
\def\cquarkbar {{\ensuremath{\overline \cquark}}\xspace}

\def\bquark    {{\ensuremath{\Pb}}\xspace}

\def\pion   {{\ensuremath{\Ppi}}\xspace}

\def\pip    {{\ensuremath{\pion^+}}\xspace}
\def\pim    {{\ensuremath{\pion^-}}\xspace}

\def\kaon    {{\ensuremath{\PK}}\xspace}

\def\KorKbar {\kern \thebaroffset\optbar{\kern -\thebaroffset \PK}{}\xspace}

\def\Kp      {{\ensuremath{\kaon^+}}\xspace}
\def\Km      {{\ensuremath{\kaon^-}}\xspace}

\newcommand{\phiz}{\ensuremath{\Pphi}\xspace}

\def\Dbar    {{\ensuremath{\offsetoverline{\PD}}}\xspace}
\def\D       {{\ensuremath{\PD}}\xspace}

\def\DorDbar {\kern \thebaroffset\optbar{\kern -\thebaroffset \PD}\xspace}
\def\Dz      {{\ensuremath{\D^0}}\xspace}

\def\Dp      {{\ensuremath{\D^+}}\xspace}
\def\Dm      {{\ensuremath{\D^-}}\xspace}

\def\DpDm    {\ensuremath{\Dp {\kern -0.16em \Dm}}\xspace}

\def\B       {{\ensuremath{\PB}}\xspace}

\def\BorBbar {\kern \thebaroffset\optbar{\kern -\thebaroffset \PB}\xspace}
\def\Bz      {{\ensuremath{\B^0}}\xspace}

\def\Bd      {{\ensuremath{\B^0}}\xspace}

\def\BdorBdbar {\kern \thebaroffset\optbar{\kern -\thebaroffset \Bd}\xspace}

\def\Bs      {{\ensuremath{\B^0_\squark}}\xspace}

\def\BsorBsbar {\kern \thebaroffset\optbar{\kern -\thebaroffset \Bs}\xspace}

\def\jpsi     {{\ensuremath{{\PJ\mskip -3mu/\mskip -2mu\Ppsi\mskip 2mu}}}\xspace}

\def\etac     {{\ensuremath{\Peta_\cquark}}\xspace}

\def\Y#1S{\ensuremath{\PUpsilon{(#1S)}}\xspace}

\def\proton      {{\ensuremath{\Pp}}\xspace}
\def\antiproton  {{\ensuremath{\overline \proton}}\xspace}

\def\Lz          {{\ensuremath{\PLambda}}\xspace}

\def\LorLbar     {\kern \thebaroffset\optbar{\kern -\thebaroffset \PLambda}\xspace}

\def\Lc          {{\ensuremath{\Lz^+_\cquark}}\xspace}

\def\Lb           {{\ensuremath{\Lz^0_\bquark}}\xspace}

\def\BF         {{\ensuremath{\mathcal{B}}}\xspace}
\def\BR         {\BF}

\newcommand{\decay}[2]{\ensuremath{#1\!\to #2}\xspace} 

\def\to                 {\ensuremath{\rightarrow}\xspace}

\def\AT#1     {\ensuremath{A_{\mathrm{T}}^{#1}}\xspace}           

\def\C#1      {\ensuremath{\mathcal{C}_{#1}}\xspace}                       
\def\Cp#1     {\ensuremath{\mathcal{C}_{#1}^{'}}\xspace}                    
\def\Ceff#1   {\ensuremath{\mathcal{C}_{#1}^{\mathrm{(eff)}}}\xspace}        
\def\Cpeff#1  {\ensuremath{\mathcal{C}_{#1}^{'\mathrm{(eff)}}}\xspace}       
\def\Ope#1    {\ensuremath{\mathcal{O}_{#1}}\xspace}                       
\def\Opep#1   {\ensuremath{\mathcal{O}_{#1}^{'}}\xspace}                    

\newcommand{\aunit}[1]{\ensuremath{\text{\,#1}}}       

\newcommand{\tev}{\aunit{Te\kern -0.1em V}\xspace}
\newcommand{\gev}{\aunit{Ge\kern -0.1em V}\xspace}
\newcommand{\mev}{\aunit{Me\kern -0.1em V}\xspace}
\newcommand{\kev}{\aunit{ke\kern -0.1em V}\xspace}
\newcommand{\ev}{\aunit{e\kern -0.1em V}\xspace}
 
\newcommand{\mevc}{\ensuremath{\aunit{Me\kern -0.1em V\!/}c}\xspace}
\newcommand{\gevc}{\ensuremath{\aunit{Ge\kern -0.1em V\!/}c}\xspace}
\newcommand{\mevcc}{\ensuremath{\aunit{Me\kern -0.1em V\!/}c^2}\xspace}
\newcommand{\gevcc}{\ensuremath{\aunit{Ge\kern -0.1em V\!/}c^2}\xspace}

\def\fb   {\ensuremath{\aunit{fb}}\xspace}
\def\invfb   {\ensuremath{\fb^{-1}}\xspace}

\newcommand{\chisq}{\ensuremath{\chi^2}\xspace}

\newcommand{\chisqip}{\ensuremath{\chi^2_{\text{IP}}}\xspace}

\def\gsim{{~\raise.15em\hbox{$>$}\kern-.85em
          \lower.35em\hbox{$\sim$}~}\xspace}
\def\lsim{{~\raise.15em\hbox{$<$}\kern-.85em
          \lower.35em\hbox{$\sim$}~}\xspace}

\def\sPlot{\mbox{\em sPlot}\xspace}

\def\sqs   {\ensuremath{\protect\sqrt{s}}\xspace}

\def\pt         {\ensuremath{p_{\mathrm{T}}}\xspace}

\def\tell1  {TELL1\xspace}
\def\ukl1   {UKL1\xspace}

\usepackage{cite} 
\usepackage{mciteplus}

\usepackage{longtable} 

\begin{document}

\renewcommand{\thefootnote}{\fnsymbol{footnote}}
\setcounter{footnote}{1}

\begin{titlepage}
\pagenumbering{roman}

\vspace*{-1.5cm}
\centerline{\large EUROPEAN ORGANIZATION FOR NUCLEAR RESEARCH (CERN)}
\vspace*{1.5cm}
\noindent
\begin{tabular*}{\linewidth}{lc@{\extracolsep{\fill}}r@{\extracolsep{0pt}}}
\ifthenelse{\boolean{pdflatex}}
{\vspace*{-1.5cm}\mbox{\!\!\!\includegraphics[width=.14\textwidth]{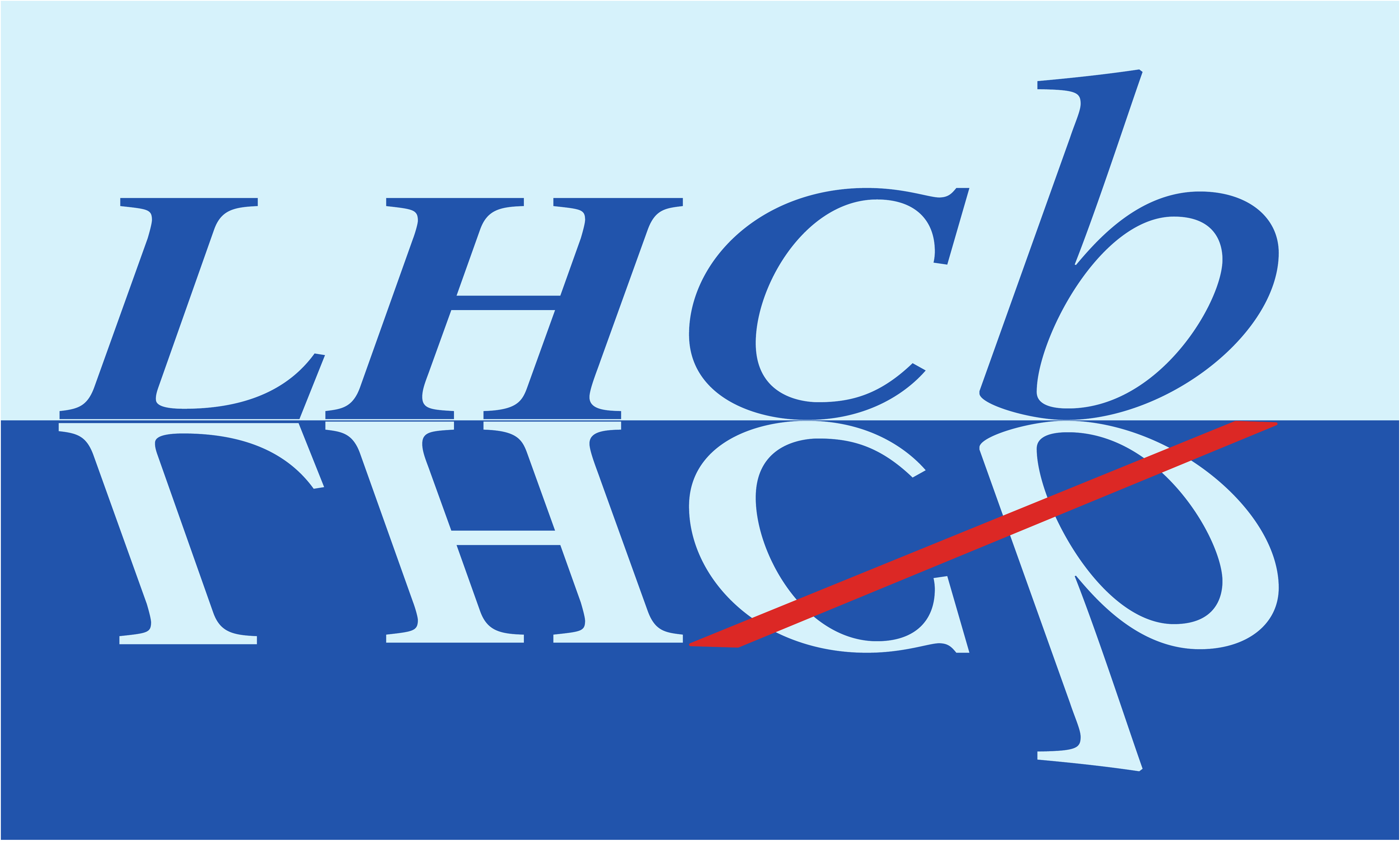}} & &}
{\vspace*{-1.2cm}\mbox{\!\!\!\includegraphics[width=.12\textwidth]{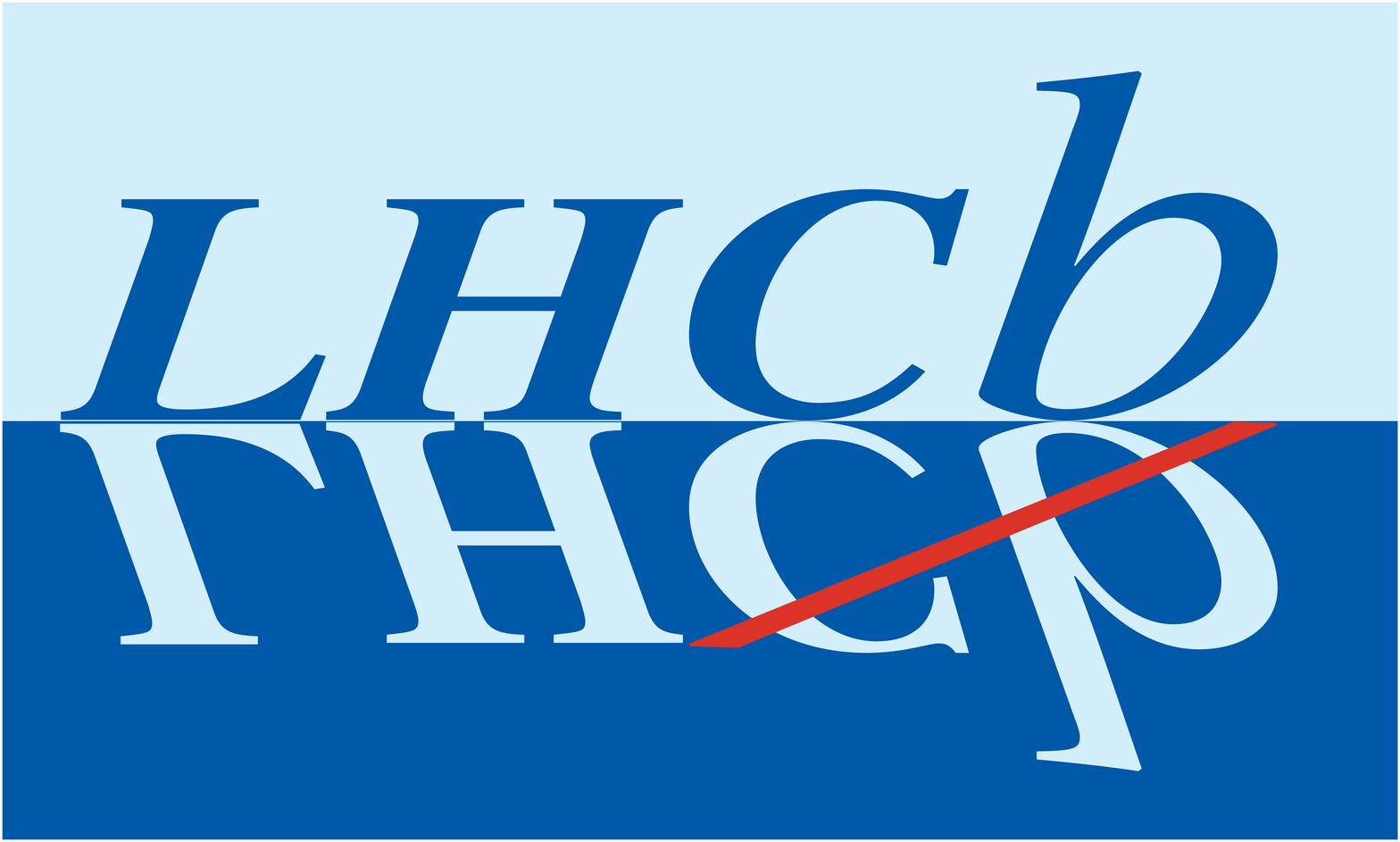}} & &}
\\
 & & CERN-EP-2020-124 \\  
 & & LHCb-PAPER-2020-012 \\  
 & & December 23, 2020 \\ 
 & & \\
\end{tabular*}

\vspace*{4.0cm}

{\normalfont\bfseries\boldmath\huge
\begin{center}
  \papertitle 
\end{center}
}

\vspace*{2.0cm}

\begin{center}
\paperauthors\footnote{Full author list given at the end of the paper.}
\end{center}

\vspace{\fill}

\begin{abstract}
  \noindent
  The decay $\decay{\Lb}{\etac(1S)\proton\Km}$ is observed for the first time using a data sample of proton-proton
  collisions, corresponding to an integrated luminosity of 5.5\invfb, collected with the LHCb experiment at
  a center-of-mass energy of 13\tev. The branching fraction of the decay is measured, using the  \decay{\Lb}{\jpsi\proton\Km} decay
  as a normalization mode,  to be $\BF(\decay{\Lb}{\etac(1S)\proton\Km})=(1.06\pm0.16\pm0.06^{+0.22}_{-0.19})\times10^{-4}$,
  where the quoted uncertainties are statistical, systematic and due to external inputs, respectively.
  A study of the $\etac(1S)\proton$ mass spectrum is performed to search for the $P_c(4312)^+$ pentaquark state. No evidence is observed and an upper limit of
  \begin{displaymath}
  \frac{\BF(\decay{\Lb}{P_c(4312)^+ \Km})\times\BF(\decay{P_c(4312)^+}{\etac(1S)\proton})}{\BF(\decay{\Lb}{\etac(1S)\proton\Km})} < 0.24
  \end{displaymath}
  is obtained at the 95\% confidence level.

\end{abstract}

\vspace*{2.0cm}

\begin{center}
  Published in Phys. Rev. D102 (2020) 112012
\end{center}

\vspace{\fill}

{\footnotesize 
\centerline{\copyright~\papercopyright. \href{\paperlicenceurl}{\paperlicence}.}}
\vspace*{2mm}

\end{titlepage}

\newpage
\setcounter{page}{2}
\mbox{~}

\renewcommand{\thefootnote}{\arabic{footnote}}
\setcounter{footnote}{0}

\cleardoublepage

\pagestyle{plain} 
\setcounter{page}{1}
\pagenumbering{arabic}

The existence of baryons comprising four quarks and an antiquark was proposed by Gell-Mann~\cite{GellMann:1964nj} 
and Zweig~\cite{Zweig:1964jf}.  Hereafter, these states are referred to as pentaquarks~\cite{Lipkin:1987sk}.  
Two pentaquark candidates were observed in the \mbox{\jpsi\proton} system of ${\decay{\Lb}{\jpsi\proton\Km}}$ decays (charge conjugation is implied throughout the text) in a data sample collected with  
the LHCb experiment during the 2011-2012 data-taking period~\cite{LHCb-PAPER-2015-029}. These candidates were labeled 
 $P_c(4450)^+$ and $P_c(4380)^+$. Using a larger data sample of  \decay{\Lb}{\jpsi\proton\Km} decays, a new pentaquark state, $P_c(4312)^+$, was  observed, and 
the broad $P_c(4450)^+$ structure resolved into two narrower  overlapping structures, labeled  
$P_c(4440)^+$ and $P_c(4457)^+$~\cite{Aaij:2019vzc}.
Many theoretical models have been proposed to describe the dynamics of the observed states, including tightly bound \dquark\uquark\uquark\cquark\cquarkbar pentaquark states \cite{Maiani:2015vwa, Lebed:2015tna, Anisovich:2015cia, Li:2015gta, Ghosh:2017fwg, Wang:2015epa, Zhu:2015bba}, baryon-meson molecular states \cite{Voloshin:2019aut, Sakai:2019qph, Wang:2019spc, Karliner:2015ina, Chen:2015loa, Chen:2015moa, Roca:2015dva, He:2015cea, Huang:2015uda}, or peaking structures due to triangle-diagram processes \cite{Guo:2015umn, Meissner:2015mza, Liu:2015fea, Mikhasenko:2015vca}. More  experimental and theoretical scrutiny is required to verify these models. 

The yet-unobserved $\decay{\Lb}{\etac\proton\Km}$ decay, where \etac refers to the $\etac(1S)$ meson, 
can provide a unique approach to search for new pentaquarks, and to study the observed states.   
It has been predicted that a $\Dbar\PSigma_{c}$ molecular state, with a mass of around 4265\mevcc, can contribute to the 
decay \mbox{\decay{\Lb}{\etac\proton\Km}} via \etac\proton final-state  interactions~\cite{Xie:2017gwc}. 
The observed $P_c(4312)^+$ state could be such a molecular state~\cite{Wu:2019adv}, since its mass is close to the $\Dbar\PSigma_{c}$ threshold~\cite{Aaij:2019vzc}.

The study of the \mbox{$\decay{\Lb}{\etac\proton\Km}$} decay provides a new way to test the binding mechanism of pentaquark states, 
as the predicted ratio of the branching fractions for a pentaquark decaying into $\etac\proton$ compared to the $\jpsi\proton$ final states depends on the pentaquark model.  The branching fraction of $\decay{P_c(4312)^+}{\etac\proton}$ is predicted to be 3 times larger than that of the \jpsi\proton decay mode if the $P_c(4312)^+$ state is a $\Dbar \PSigma_c$ molecule~\cite{Voloshin:2019aut, Sakai:2019qph, Wang:2019spc}.

This paper presents the first observation of the \decay{\Lb}{\etac\proton\Km} decay, with the \etac meson 
reconstructed using the \decay{\etac}{\proton\antiproton} decay mode, and reports a search for the $P_c(4312)^+$ pentaquark state in the $\etac\proton$ system. 
The analysis uses the decay \decay{\Lb}{\jpsi\proton\Km} as a normalization channel, where the \jpsi meson decays to  \proton\antiproton.   
The data sample used in this analysis corresponds to an integrated luminosity of 5.5\invfb, collected with the LHCb experiment in proton-proton collisions at $\sqs=$13\tev between 2016 and 2018.

In the \B-meson sector, heavy quark effective theory  \cite{Deshpande:1994mk, Ahmady:1994vq} predicts that the decay rates of the ${\decay{B}{\etac X}}$ and ${\decay{B}{\jpsi X}}$ channels are of the same order of magnitude.  Experimental results are in good agreement with this expectation~\cite{PDG2019}. 
Studying the branching fraction ratio between the \mbox{\decay{\Lb}{\etac\proton\Km}} and  \mbox{\decay{\Lb}{\jpsi\proton\Km}} decays will provide the first comparison of \bquark-baryon decay rates to the $\etac X$ and $\jpsi X$ final states, and help to test whether the presence of an additional spectator quark modifies the final-state interactions in a non-negligible way.

The LHCb detector is a single-arm forward spectrometer covering the pseudorapidity range $2 < \eta < 5$, and is described in detail in Refs.~\cite{Alves:2008zz, Aaij:2014jba}. The detector includes a silicon-strip vertex detector surrounding the proton-proton interaction region, tracking stations on either side of a dipole magnet, ring-imaging Cherenkov (RICH) detectors, calorimeters and muon chambers. The online event selection is performed by a trigger~\cite{LHCb-DP-2019-001}, which consists of a hardware stage, based on information from the calorimeter and muon systems, followed by a software stage, which applies a full event reconstruction. The software trigger requires a two-, three- or four-track secondary vertex  with a significant displacement from any primary vertex (PV) that is consistent with originating from the decay of a  \bquark hadron~\cite{Gligorov:2012qt}.

Simulated data samples as described in Refs.~\cite{Sjostrand:2007gs, LHCb-PROC-2010-056, Lange:2001uf, Golonka:2005pn, Allison:2006ve, LHCb-PROC-2011-006},
are used to optimize the event selection, determine the efficiency of the reconstruction and event selection, and to constrain the fit model which determines  the signal yield.
The simulated \mbox{$\decay{\Lb}{\etac\proton\Km}$} and 
$\decay{\Lb}{\jpsi\proton\Km}$ decays are generated based on a uniform phase-space model.
The simulated decays are also weighted to match the \Lb momentum spectrum and Dalitz-plot distribution in the data, as described later in this paper.

The $\decay{\Lb}{\etac(\to\proton\antiproton)\proton\Km}$, and 
$\decay{\Lb}{\jpsi(\to\proton\antiproton)\proton\Km}$ candidates are reconstructed and selected using the same selection criteria, with a $\proton\antiproton$ mass window 
of $[2800,3200]\mevcc$ that covers both the \etac and \jpsi mass regions.    
In the following, the notation $[\cquark\cquarkbar]$ will be used to 
refer to both the \etac and the \jpsi candidates from \Lb baryon decays.
Particle identification (PID) variables in the simulation are calibrated using large data samples 
of kinematically identified protons and kaons, originating from $\decay{\Lb}{\Lc(\to\proton\Km\pip)\pim}$ and 
\mbox{\decay{\Dz}{\Km\pip}} decays.

The offline event selection is performed using a preselection, followed by a requirement on the 
response of a boosted decision tree (BDT) classifier~\cite{FREUND1997119, Hocker:2007ht}. 
In the preselection, each track is required to be of good quality. Kaons and protons are both required to have $\pt>300\mevc$, where $\pt$ is the component of the momentum 
transverse to the beam. Protons are also required to have a momentum larger than 10\gevcc, such that the kaons and protons can be distinguished by the RICH detectors. The sum of the \pt of the proton and kaon from the \Lb baryon is required to be larger than 900\mevc. 
The $[\cquark\cquarkbar]$ candidate is required to have a good-quality vertex. 

The \Lb candidate must have a good-quality decay vertex that is significantly displaced from 
every PV, and have \mbox{$\chisqip < 25$} with respect to the associated PV. Here, $\chisqip$ is defined as the \chisq difference 
between the vertex fit of a PV reconstructed with or without the particle in question, and the associated PV 
is the one with the smallest \chisqip value. The angle between the 
reconstructed momentum vector of the \Lb candidate and the line connecting the associated PV and the \Lb decay 
vertex, $\theta_\Lb$, is required to satisfy $\cos(\theta_\Lb)>0.9999$. 

Contamination from 
\mbox{\decay{\Bs}{\proton\antiproton\Kp\Km}} and \mbox{\decay{\Bz}{\proton\antiproton\Kp\pim}} decays, 
where a kaon or pion is misidentified as a proton, is removed by applying strict particle 
identification requirements on candidates with a mass within $\pm 50 \mevcc$ around the known \Bs or \Bz mass~\cite{PDG2019} after assigning a kaon or pion mass hypothesis to the proton. Backgrounds from $\decay{\phiz(1020)}{\Kp\Km}$ and $\decay{\Dz}{\Kp\Km}$ decays, where one of the kaons is misidentified as a proton and the \Lb candidate is formed by combining the particles with a $[\cquark\cquarkbar]$ candidate from elsewhere in the event, are also observed.
These contributions are  
removed by placing stricter particle-identification requirements on 
candidates with a $\proton\Km$ mass within $\pm 10 \mevcc$ ($\pm 20 \mevcc$) of the 
known $\phiz(1020)$ ($\Dz$) mass,  after 
assigning a kaon mass hypothesis~\cite{PDG2019} to the proton.

\begin{figure}[t]
  \begin{center}
   \includegraphics[width=0.49\textwidth]{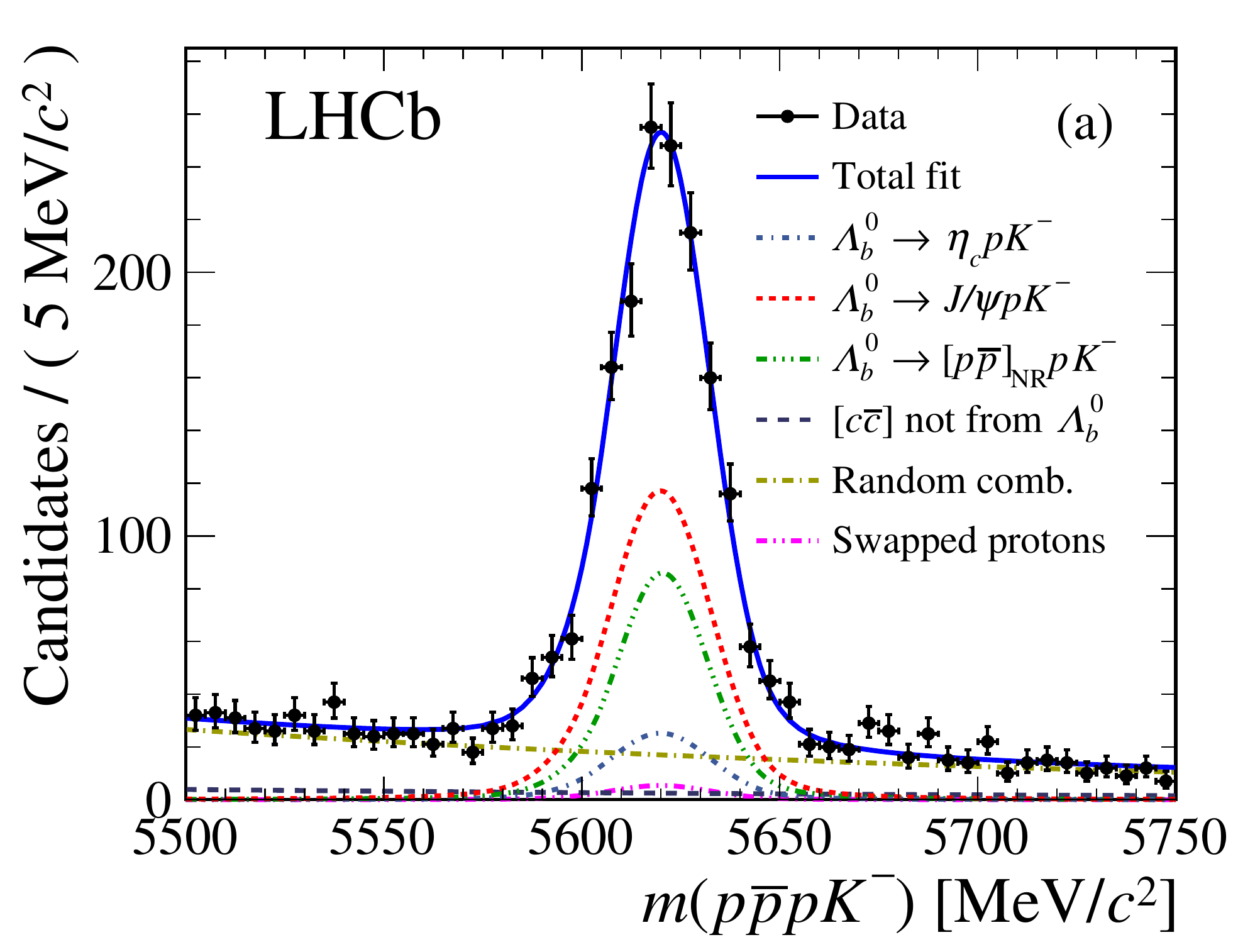}
    \includegraphics[width=0.49\textwidth]{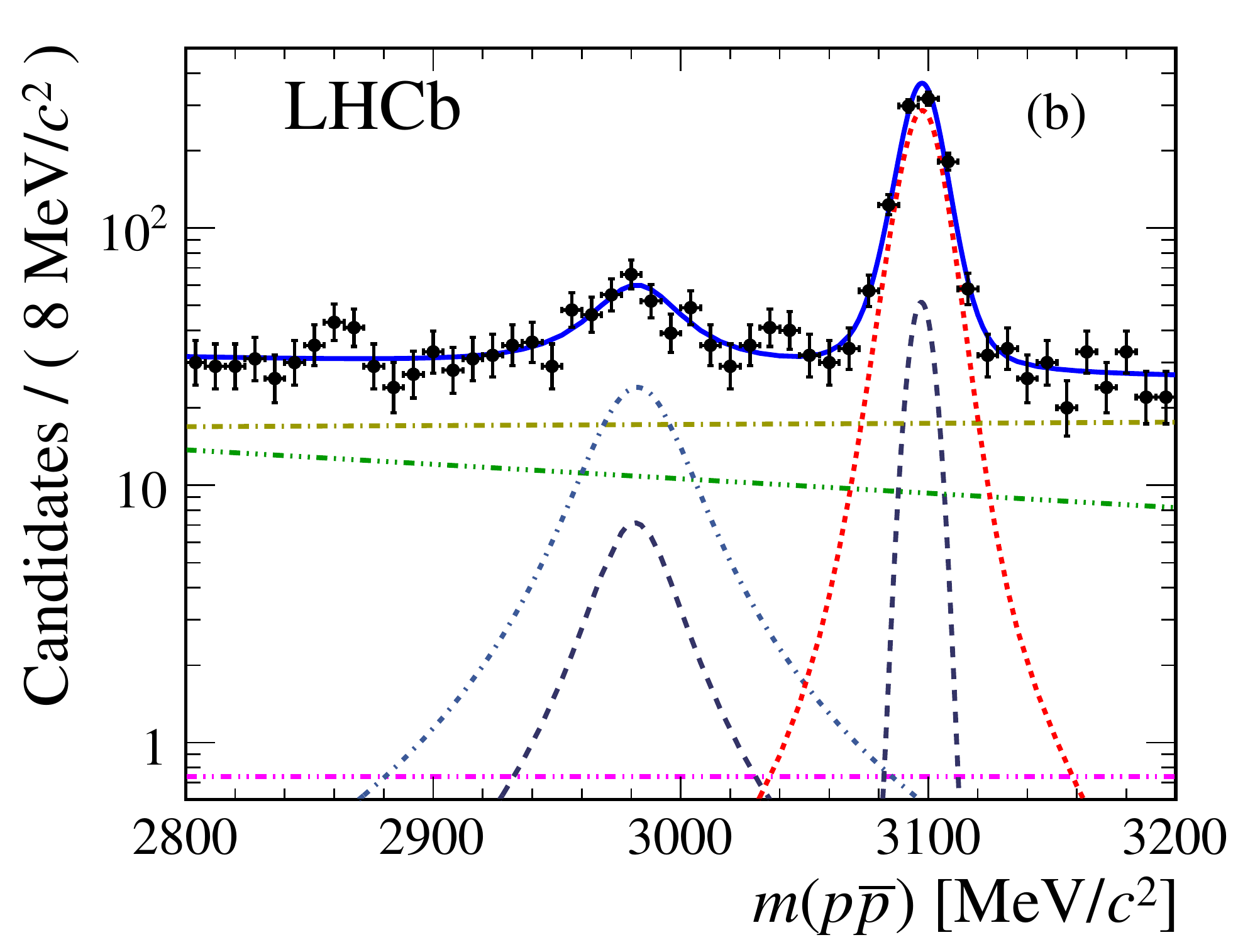}\\
  \end{center}
  \caption{Distributions of  (a) $m(\proton\antiproton\proton\Km)$ 
and (b) $m(\proton\antiproton)$   of the selected candidates. 
 The data are shown as black circles, while the blue solid line shows the fit result. Individual components are given in the legend. 
} 
  \label{fig:nominal_fit}
\end{figure}

After the preselection,  further separation between the signal and combinatorial backgrounds originating from a random combination of final-state particles is 
achieved by using a BDT classifier. The classifier uses the following input variables:  the \pt of the \Lb candidate, and of the kaon and proton directly from the \Lb decay; 
the \chisqip of the \Lb candidate, the $[\cquark\cquarkbar]$ candidate, and the kaon and proton directly from the \Lb decay;  
the smallest values of both the \pt and \chisqip of the $[\cquark\cquarkbar]$ decay products; 
the significance of 
the displacement of the \Lb vertex with respect to the associated PV; the vertex-fit $\chisq$ 
of the \Lb candidate; the $\theta_\Lb$ angle; and the PID information of the final-state particles. 
The BDT is trained using simulated \mbox{\decay{\Lb}{\etac\proton\Km}} decays for the signal, and the data candidates 
in the \mbox{\proton\antiproton\proton\Km} invariant-mass sideband above 5800\mevcc for the background. The  
requirement on the BDT response is optimized by maximizing the figure of merit $\epsilon^{\rm sig}/(a/2+\sqrt{N_{\rm bkg}})$~\cite{Punzi:2003bu}, 
where $\epsilon^{\rm sig}$ is the BDT selection efficiency estimated using the simulated \mbox{\decay{\Lb}{\etac\proton\Km}} sample, $a=5$ 
is the target significance for the signal in standard deviations, and $N_{\rm bkg}$ is the expected yield of background with $\proton\antiproton$ and $\proton\antiproton\proton\Km$ masses in the ranges $m(\proton\antiproton)\in[2951.4, 3015.4]\mevcc$ and $m(\proton\antiproton\proton\Km)\in[5585, 5655]\mevcc$, respectively. The background yields are estimated using the \mbox{\proton\antiproton\proton\Km} and \mbox{\proton\antiproton} invariant-mass sidebands in the data. 
The BDT response requirement provides about 70\% signal efficiency  
and suppresses the background by a factor of approximately 100.
After the BDT selection, a background in the normalization channel is observed due to swapping the proton from the \Lb decay with a proton from the \jpsi decay. This contribution is removed by requiring the invariant mass of the system formed by the proton from the \Lb baryon and the antiproton from the \jpsi meson to be inconsistent with the known \jpsi mass~\cite{PDG2019}.  The \mbox{\proton\antiproton\proton\Km} and 
\mbox{\proton\antiproton} invariant-mass spectra of the selected data are displayed in Fig.~\ref{fig:nominal_fit}.

A two-dimensional unbinned maximum-likelihood fit to the   \mbox{\proton\antiproton\proton\Km} and 
\mbox{\proton\antiproton} invariant-mass distributions is performed to determine  the signal yield.  
The \mbox{\proton\antiproton\proton\Km} mass spectra of the signal and normalization channels are described 
using the same model, sharing the shape parameters. The signal is modeled by the sum of two Crystal Ball (CB) functions~\cite{Skwarnicki:1986xj} 
with common peak positions. 
The tail parameters of the CB functions are determined from simulation, while the mean and width of the Gaussian cores are freely varying in the fit to the data. 
The \mbox{\proton\antiproton} mass spectrum is described with a relativistic Breit--Wigner function~\cite{Jackson:1964zd} 
convolved with a Gaussian resolution function for the $\etac$, and is described with the sum of two CB functions with common peak positions for the $\jpsi$ decay.

When modeling the $m(\proton\antiproton)$ spectrum, the correlation between $m(\proton\antiproton\proton\Km)$ and $m(\proton\antiproton)$ needs to be taken into account. 
The width (peak) parameter of the resolution function of the signal channel, and the width (peak) parameters of the Gaussian cores for the normalization channel, are 
parameterized as second-order (first-order) polynomial functions of $m(\proton\antiproton\proton\Km)$; the coefficients  
of these polynomial functions are calibrated using simulated samples.

For the two-dimensional mass spectrum of the background components, 
it is assumed that $m(\proton\antiproton\proton\Km)$ and $m(\proton\antiproton)$ are uncorrelated, which is 
corroborated using the background-dominated data sample before the BDT selection is applied. For background  
from \mbox{\decay{\Lb}{\proton\antiproton\proton\Km}} decays but with the \proton\antiproton pair not originating from a  \etac or \jpsi resonance, 
the 
$m(\proton\antiproton)$ spectrum is described using an exponential function, and
the $m(\proton\antiproton\proton\Km)$ 
spectrum is described 
using the same model as the signal but the parameters of the distribution are allowed to take different values in the fit. 
For background with a 
$\decay{[\cquark\cquarkbar]}{\proton\antiproton}$ process but not from a \Lb decay, the $m(\proton\antiproton\proton\Km)$ 
distribution is described using an exponential function, and the $m(\proton\antiproton)$ spectrum is modeled by 
Breit--Wigner functions that are each convolved with a separate Gaussian function to describe the \etac and \jpsi resonances. 
In the fit, a Gaussian constraint of $31.9 \pm 0.7 \mevcc$ ~\cite{PDG2019} is applied to the natural width of the  \etac meson for both the signal and background components.  For combinatorial backgrounds, both the 
$m(\proton\antiproton\proton\Km)$ and $m(\proton\antiproton)$ spectra are described using exponential functions.  
The background shape due to swapping the two protons in the $\decay{\Lb}{\etac(\to\proton\antiproton)\proton\Km}$ decay shares the
same shape in $m(\proton\antiproton\proton\Km)$ as the signal channel, 
while the $m(\proton\antiproton)$ shape, 
and the relative yield with respect to the signal component of the signal channel, are determined from simulation.  
Given the limited yield of 
 \mbox{\decay{\Lb}{\etac\proton\Km}} decays expected in this data sample, 
the interference between the \mbox{\decay{\Lb}{\etac\proton\Km}} 
and nonresonant \mbox{\decay{\Lb}{\proton\antiproton\proton\Km}} decays is not considered. 
An amplitude analysis of a larger data set is needed to have sensitivity to such interference effects. 

The $m(\proton\antiproton\proton\Km)$ and $m(\proton\antiproton)$ distributions of the selected candidates are presented in Fig.~\ref{fig:nominal_fit}, 
with the one-dimensional projections of the fit overlaid.
The yields of the signal and normalization modes are 
$N(\decay{\Lb}{\etac\proton\Km}) = 173\pm25$ and  $N(\decay{\Lb}{\jpsi\proton\Km}) = 804\pm31$, respectively, where the uncertainties are statistical only.  
To estimate the signal significance, a two-dimensional fit without the contribution from the \mbox{\decay{\Lb}{\etac\proton\Km}} decay is performed. The difference in log-likelihood between this and the nominal fit is found to be 29.4. Based on the assumption of a \chisq distribution with one degree of freedom, the statistical significance of the \mbox{\decay{\Lb}{\etac\proton\Km}} decay with respect to the background-only hypothesis, expressed in Gaussian standard deviations, is $7.7 \sigma$.

The ratio of the branching fraction between the \mbox{\decay{\Lb}{\etac\proton\Km}} and \mbox{\decay{\Lb}{\jpsi\proton\Km}} decays is given by 
\begin{equation}
\begin{split}
\label{eq:master_formula}
\frac{\BF(\decay{\Lb}{\etac\proton\Km})}{\BF(\decay{\Lb}{\jpsi\proton\Km})} =  
 \frac{N(\decay{\Lb}{\etac\proton\Km})}{N(\decay{\Lb}{\jpsi\proton\Km})} \times \frac{\epsilon(\decay{\Lb}{\jpsi\proton\Km})}{\epsilon(\decay{\Lb}{\etac\proton\Km})} \times \frac{\BF(\decay{\jpsi}{\proton\antiproton})}{\BF(\decay{\etac}{\proton\antiproton})},
\end{split}
\end{equation}
where $N$ represents the yield of the decay given in the parentheses, 
determined from a fit to the invariant-mass spectrum
and $\epsilon$ is the  efficiency accounting for 
the detector geometrical acceptance, reconstruction and event selection. 
The known values of the branching fractions, $\BF$, 
of the \mbox{\decay{\Lb}{\jpsi\proton\Km}}, \mbox{\decay{\jpsi}{\proton\antiproton}}~\cite{PDG2019} and \mbox{\decay{\etac}{\proton\antiproton}} decays~\cite{PDG2020} are used as external inputs for the measurement of \BF(\decay{\Lb}{\etac\proton\Km}).

The efficiencies of the  detector geometrical  acceptance, reconstruction and event selections  are determined from simulation. The 
agreement between data and simulation is improved by weighting the two-dimensional ($p,\pt$) distribution of the \Lb baryons in simulation. The weights are obtained using a comparison between a large sample of data 
and simulated events from  \mbox{\decay{\Lb}{\jpsi\proton\Km}} decays, where the \jpsi meson is reconstructed through its decay \mbox{\decay{\jpsi}{\mup\mun}}. The distributions of $m(\proton\Km)$ and 
$m([\cquark\cquarkbar]\proton)$ in the simulation for signal and normalization channels are also weighted to 
match the corresponding distributions observed in data, where the data distributions are obtained using the \sPlot 
technique~\cite{Pivk:2004ty} with $m(\proton\antiproton\proton\Km)$ and $m(\proton\antiproton)$ as the 
discriminating variables.  The ratio between the overall efficiencies of the signal and normalization channels is 
$0.95\pm0.02$, where the uncertainty accounts only for the finite yields  of the simulated events.  The ratio of branching fractions between the \mbox{\decay{\Lb}{\etac\proton\Km}} and \mbox{\decay{\Lb}{\jpsi\proton\Km}} decays is obtained as
\begin{align*}
\frac{\BF(\decay{\Lb}{\etac\proton\Km})}{\BF(\decay{\Lb}{\jpsi\proton\Km})} = 0.333 \pm 0.050, 
\end{align*}
\noindent where the quoted uncertainty is statistical only.  

\begin{figure}[t]
  \begin{center}
    \includegraphics[width=0.47\textwidth]{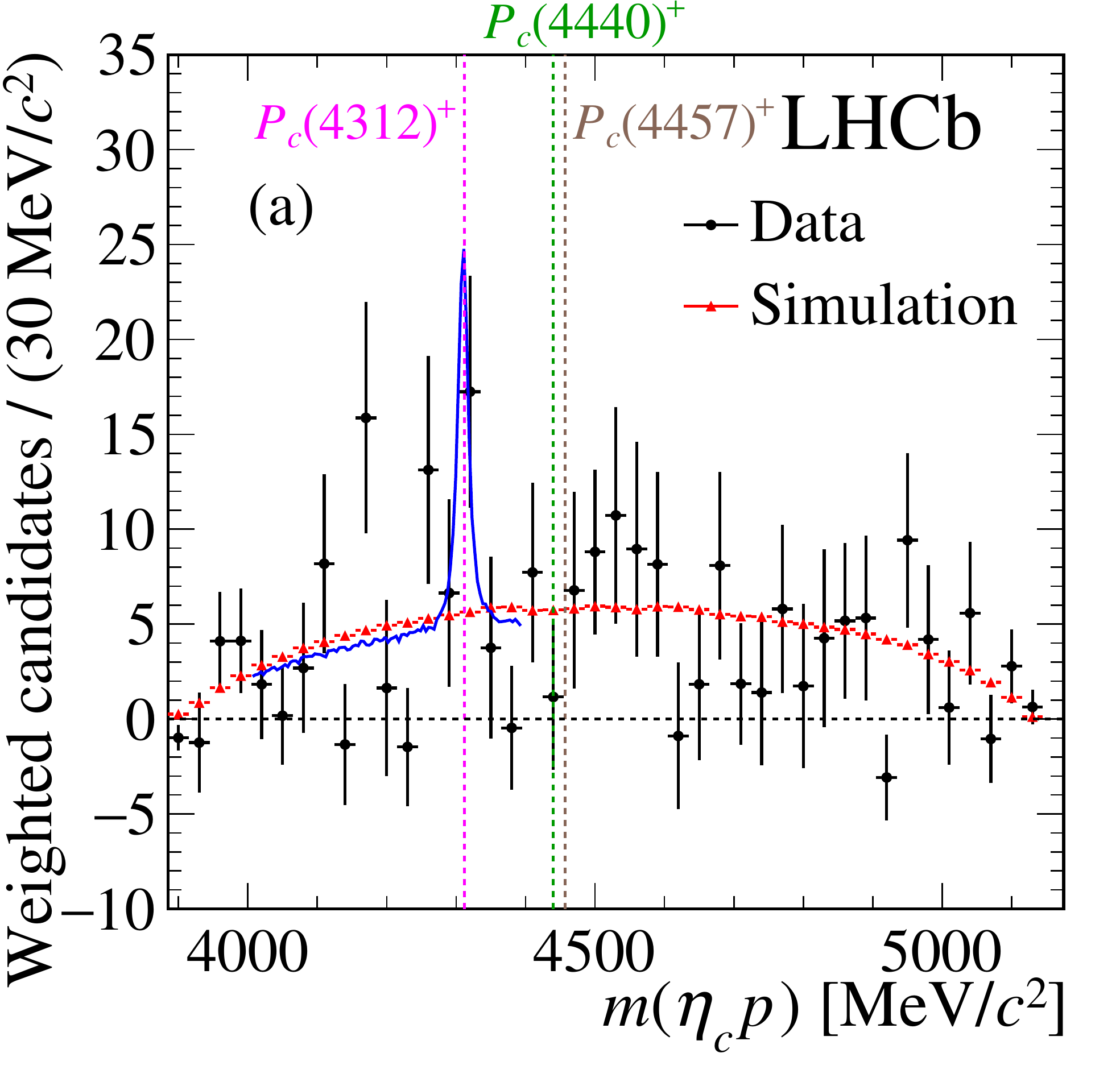}
    \includegraphics[width=0.47\textwidth]{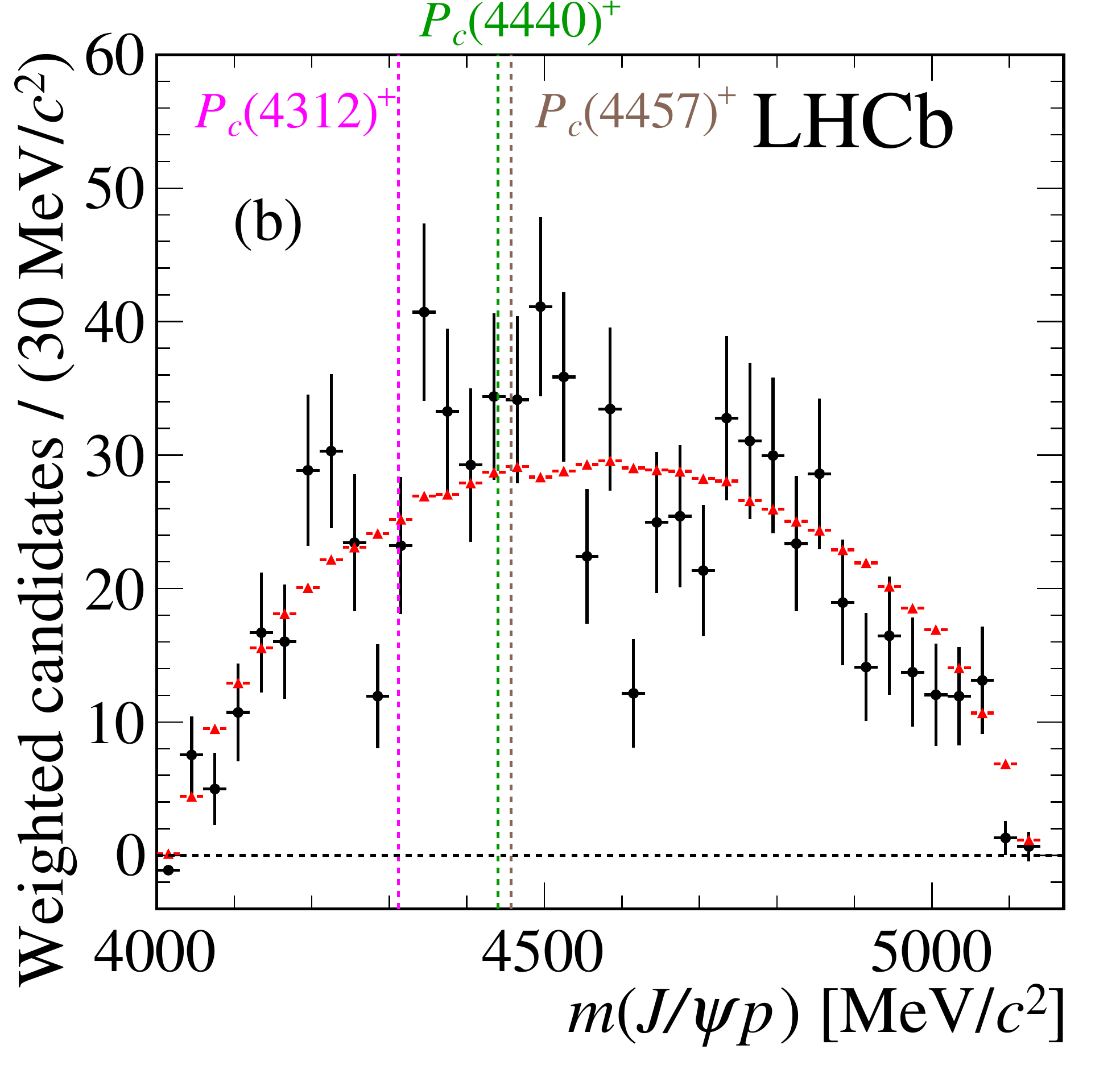}
  \end{center}
  \caption{The invariant-mass spectra of (a) the \mbox{\etac\proton} system of the  
\mbox{\decay{\Lb}{\etac\proton\Km}} decays and (b) the \mbox{\jpsi\proton} system of the \mbox{\decay{\Lb}{\jpsi\proton\Km}} decays. The black points represent the background-subtracted data and the red points correspond to the expectation from a simulation generated according to a uniform phase-space model. The blue solid line in (a) shows the fit projection of the \etac\proton mass spectrum including the contribution from a $P_c(4312)^+$ resonance in the mass range $[4000, 4400]\mevcc$.}
  \label{fig:etacp}
\end{figure}

A search for a $P_c(4312)^+\to\etac\proton$ contribution to the $\decay{\Lb}{\etac\proton\Km}$ decay is performed by projecting out the background-subtracted $\etac\proton$ mass spectrum using the \sPlot technique. The resulting $\etac\proton$ (and $\jpsi\proton$) mass distributions are shown in Fig.~\ref{fig:etacp}. 
A weighted unbinned maximum-likelihood fit~\cite{Xie:2009rka} is applied to the \etac\proton mass spectrum, where the data is described as the incoherent sum of $P_c(4312)^+\to\etac\proton$ decays and a nonresonant $\etac\proton$ contribution. The $P_c(4312)^+$ resonance is modeled using a relativistic Breit--Wigner function~\cite{Jackson:1964zd}, with parameters obtained from Ref.~\cite{Aaij:2019vzc}, and is convolved with the sum of two Gaussian resolution functions whose shape parameters are determined from simulation. The contribution from \decay{\Lb}{\etac\proton\Km} decays with a  non-resonant \etac\proton system is modeled using simulated events generated with a uniform phase-space model. The fit projection is shown in Fig.~\ref{fig:etacp}\,(a). 

The yield of the $P_c(4312)^+$ state is determined to be $16^{+12}_{-~9}\,({\rm stat.}) \pm 4\,({\rm syst.})$. The systematic uncertainty on the yield is estimated by using alternative models to describe the \Lb component without \etac\proton resonances, and varying the mass and width of the $P_c(4312)^+$ state based on their uncertainties from Ref.~\cite{Aaij:2019vzc}. To consider the potential influence of the interference between the $P_c(4312)^+$ component and reflections from $\PLambda^*\to\proton\Km$ resonances, several $\decay{\Lb}{\jpsi\proton\Km}$ samples are generated based on the result of a full amplitude fit to the  \mbox{$\decay{\Lb}{\jpsi(\to\mup\mun)\proton\Km}$} sample used in Ref.~\cite{Aaij:2019vzc}, with a different scale factor assigned on the $P_c(4312)^+$ amplitude to account for a change in its contribution. 
A fit is performed to these simulated $\jpsi\proton$ mass spectra, using the same description for the $P_c(4312)^+$ contribution  as that in the fit model of the background-subtracted $\etac\proton$ mass spectrum. The largest relative difference between the $P_c(4312)^+$ relative contribution obtained from the fit and its true value in the simulated samples is taken as a systematic uncertainty for this potential interference.

The difference of the log-likelihood between the nominal fit and a fit with the $P_c(4312)^+$ yield  fixed to zero is 2.4. Since all of the shape parameters of the $P_c(4312)^+$ component are fixed in the nominal fit, the statistical significance of the $P_c(4312)^+$ state is 
2.2\,$\sigma$. 
Defining the relative $P_c(4312)^+$ contribution analogous to that which is used in Ref.~\cite{Aaij:2019vzc} as 
\begin{align}
\mathcal{R} \equiv \frac{\BF(\Lb\to P_c(4312)^+\Km)}{\BR(\Lb\to\etac\proton\Km)}\BR(P_c(4312)^+\to\etac\proton), 
\end{align}
\noindent a 95\% confidence level upper limit of $R<0.24$ is obtained from the likelihood profile distribution. The search to the $P_c(4440)^+$ and $P_c(4457)^+$ states is not performed in this paper, as they will together perform like a broad structure under the limited sample size~\cite{LHCb-PAPER-2015-029}, which cannot be disentangled from the reflections from the $\decay{\Lb}{\PLambda^*\etac}$, $\decay{\PLambda^*}{\proton\Km}$ decay chain without a full amplitude analysis.

Sources of systematic uncertainty on the \mbox{\decay{\Lb}{\etac\proton\Km}} branching fraction arise from the fitting procedure 
and limited 
knowledge of the efficiencies, and are summarized in Table~\ref{tab:systematic_summary}. 
Pseudoexperiments are used to estimate the 
effects due to parameters determined from simulation. Systematic uncertainties on the fit model are evaluated by 
using alternative fit models where: 
the exponential functions 
are replaced by Chebyshev polynomials; the contributions from genuine \Lb decays in the 
 $m(\proton\antiproton\proton\Km)$ spectrum are  modeled by the Hypatia distribution~\cite{Santos:2013gra}; the 
resolution of the \etac peaking structure in the  $m(\proton\antiproton)$ spectrum is replaced by the average resolution 
of the CB functions describing the \jpsi peak; the shape parameters of the \Lb peak in the \mbox{\decay{\Lb}{\proton\antiproton\proton\Km}} decay without 
the \etac or \jpsi resonances are fixed to be the same as those of the signal and the normalization decays. 
Pseudoexperiments are used to estimate the potential 
bias of the fit yields, which is found to be negligible compared to the statistical uncertainties.  
Based on each alternative fit model described above, the significance of the \decay{\Lb}{\etac\proton\Km} is reestimated.  The smallest significance found is  approximately $7.7 \sigma$. This is the first observation of this decay mode.  

\begin{table}[b]
\centering
\caption{Summary of the uncertainties on the branching fraction ratio  $\BF(\decay{\Lb}{\etac\proton\Km})/\BF(\decay{\Lb}{\jpsi\proton\Km})$. The total systematic uncertainty is obtained by summing the individual contributions in quadrature.}
\label{tab:systematic_summary}
\begin{tabular}{lP{3.5}}
\toprule
Source & \multicolumn{1}{c}{Uncertainty} (\%) \\ \midrule
\Lb $p$ and $\pt$ distributions &  1.0\\
$m(\proton\Km)$ and $m([\cquark\cquarkbar]\proton)$ distributions &  3.2\\
Fit model &  4.0\\
Finite simulated sample sizes & 2.5 \\
\midrule
Total systematic uncertainty &  5.8 \\
Statistical uncertainty  & 13.6 \\
$\BF(\decay{[\cquark\cquarkbar]}{\proton\antiproton})$  & 9.6 \\
\bottomrule
\end{tabular}
\end{table}

Uncertainties on the efficiency ratio between the signal and normalization channels are largely canceled due to the 
similarity of these two decay modes.  
For the estimation of systematic uncertainties related to the weighting procedure of $m([\cquark\cquarkbar]\proton)$, $m(\proton\Km)$ 
and ($p,\pt$) of the \Lb decays in simulation, pseudoexperiments are used to propagate the uncertainties of single-event weights, originating 
from the finite yield of the samples used to obtain the weights, to the uncertainty of the overall efficiency ratio; an alternative 
binning scheme is used to estimate the uncertainty 
due to the choice of binning in the weighting procedure; and the negative weights, given by the \sPlot technique due to statistical fluctuations, are set to zero to recalculate the overall efficiency ratio. A systematic uncertainty 
is also assigned for the finite size of the simulated samples used for the efficiency estimation. 

The total systematic uncertainty of the \mbox{\decay{\Lb}{\etac\proton\Km}} branching fraction measurement is obtained by 
adding the above contributions in quadrature, leading to a value of 5.8\%, and details are given in Table~\ref{tab:systematic_summary}.   The dominant contribution is the uncertainty related to the fit model. The limited knowledge of the branching fractions of the \mbox{\decay{\Lb}{\jpsi\proton\Km}}, \mbox{\decay{\jpsi}{\proton\antiproton}} and \mbox{\decay{\etac}{\proton\antiproton}} decays~\cite{PDG2019} is also considered as an external source that contributes to the total uncertainty.

The background-subtracted data distributions of $m([\cquark\cquarkbar] p)$ for the signal and normalization channels are shown 
in Fig.~\ref{fig:etacp}, with the distributions of simulated events overlaid. The background subtraction is based on the \sPlot 
technique~\cite{Pivk:2004ty}, with $m(\proton\antiproton\proton\Km)$ and $m(\proton\antiproton)$ as the
discriminating variables. No significant peaking structures are seen. 
The fractions of the $P_c(4312)^+$, $P_c(4440)^+$ and $P_c(4457)^+$ contributions to the $\decay{\Lb}{\jpsi\proton\Km}$ decays are only roughly 0.3\%, 1.1\% and 0.5\%, respectively~\cite{Aaij:2019vzc}, and given the limited  $\decay{\Lb}{\jpsi\proton\Km}$ yields of this analysis, it is not surprising that these $P_c$ contributions are not observed.

In summary, the first observation of the decay \mbox{\decay{\Lb}{\etac\proton\Km}} has been reported using proton-proton collision data  
collected with the LHCb experiment, corresponding to an integrated luminosity of 5.5\invfb. The significance of this 
observation, over the background-only hypothesis, is 7.7  standard deviations. 
The branching fraction ratio between the \mbox{\decay{\Lb}{\etac\proton\Km}} and \mbox{\decay{\Lb}{\jpsi\proton\Km}} decays is measured to be 
\begin{align*}
\frac{\BF(\decay{\Lb}{\etac\proton\Km})}{\BF(\decay{\Lb}{\jpsi\proton\Km})} = 0.333 \pm 0.050 \ ({\rm stat.}) \pm 0.019 \ ({\rm syst.}) \pm 0.032 \ (\BF)~, 
\end{align*} 
where the first uncertainty is statistical, the second is systematic, 
and the last is due to the  uncertainty on the branching 
fractions of the \mbox{\decay{\etac}{\proton\antiproton}} and \mbox{\decay{\jpsi}{\proton\antiproton}} decays. 
Using this ratio, the branching fraction of the \mbox{\decay{\Lb}{\etac\proton\Km}} decay is determined to be 
\begin{align*}
\BF(\decay{\Lb}{\etac\proton\Km}) = (1.06 \pm 0.16 \ ({\rm stat.}) \pm 0.06 \ ({\rm syst.})^{+ 0.22}_{- 0.19} \ (\BF))\times 10^{-4} , 
\end{align*}
where the third uncertainty also depends on the branching fraction of the 
\mbox{\decay{\Lb}{\jpsi\proton\Km}} decay.

The observation of this decay opens up a new line of investigation in searching for pentaquarks 
in the \mbox{\etac\proton} system. 
If the $P_c(4312)^+$ state is a $\Dbar \PSigma_c$ molecule and the predictions of Refs.~\cite{Voloshin:2019aut, Sakai:2019qph, Wang:2019spc} are accurate, a value of
$R_{\Dbar\PSigma_c}\sim0.03$ would be expected, based on the $P_c(4312)^+$ relative contribution in \mbox{\decay{\Lb}{\jpsi\proton\Km}} decays~\cite{Aaij:2019vzc} and the above result for $\BF(\decay{\Lb}{\etac\proton\Km})/\BF(\decay{\Lb}{\jpsi\proton\Km})$. The 95\% confidence level upper limit obtained in this analysis, $R<0.24$, does not exclude this molecular interpretation for the $P_c(4312)^+$ state.
A further amplitude analysis with a larger data sample is required for a more quantitative comparison to theoretical predictions~\cite{Voloshin:2019aut, Sakai:2019qph, Wang:2019spc}.  
By using an upgraded LHCb detector with improved trigger conditions and larger data samples collected, there are good prospects for using this decay to shed light on the binding mechanism of the recently observed pentaquark states~\cite{Aaij:2019vzc}.

\newpage
\section*{Acknowledgements}
\noindent We express our gratitude to our colleagues in the CERN
accelerator departments for the excellent performance of the LHC. We
thank the technical and administrative staff at the LHCb
institutes.
We acknowledge support from CERN and from the national agencies:
CAPES, CNPq, FAPERJ and FINEP (Brazil); 
MOST and NSFC (China); 
CNRS/IN2P3 (France); 
BMBF, DFG and MPG (Germany); 
INFN (Italy); 
NWO (Netherlands); 
MNiSW and NCN (Poland); 
MEN/IFA (Romania); 
MSHE (Russia); 
MICINN (Spain); 
SNSF and SER (Switzerland); 
NASU (Ukraine); 
STFC (United Kingdom); 
DOE NP and NSF (USA).
We acknowledge the computing resources that are provided by CERN, IN2P3
(France), KIT and DESY (Germany), INFN (Italy), SURF (Netherlands),
PIC (Spain), GridPP (United Kingdom), RRCKI and Yandex
LLC (Russia), CSCS (Switzerland), IFIN-HH (Romania), CBPF (Brazil),
PL-GRID (Poland) and OSC (USA).
We are indebted to the communities behind the multiple open-source
software packages on which we depend.
Individual groups or members have received support from
AvH Foundation (Germany);
EPLANET, Marie Sk\l{}odowska-Curie Actions and ERC (European Union);
A*MIDEX, ANR, Labex P2IO and OCEVU, and R\'{e}gion Auvergne-Rh\^{o}ne-Alpes (France);
Key Research Program of Frontier Sciences of CAS, CAS PIFI,
Thousand Talents Program, and Sci. \& Tech. Program of Guangzhou (China);
RFBR, RSF and Yandex LLC (Russia);
GVA, XuntaGal and GENCAT (Spain);
the Royal Society
and the Leverhulme Trust (United Kingdom).

\newpage
\addcontentsline{toc}{section}{References}
\bibliographystyle{LHCb}
\bibliography{main,standard,LHCb-PAPER,LHCb-CONF,LHCb-DP,LHCb-TDR}

\newpage

\clearpage
\centerline
{\large\bf LHCb collaboration}
\begin
{flushleft}
\small
R.~Aaij$^{31}$,
C.~Abell{\'a}n~Beteta$^{49}$,
T.~Ackernley$^{59}$,
B.~Adeva$^{45}$,
M.~Adinolfi$^{53}$,
H.~Afsharnia$^{9}$,
C.A.~Aidala$^{83}$,
S.~Aiola$^{25}$,
Z.~Ajaltouni$^{9}$,
S.~Akar$^{64}$
% ...

{\footnotesize \it

% ...
$ ^{9}$Universit{\'e} Clermont Auvergne, CNRS/IN2P3, LPC, Clermont-Ferrand, France\\
% ...
$ ^{25}$INFN Sezione di Milano, Milano, Italy\\
% ...
$ ^{31}$Nikhef National Institute for Subatomic Physics, Amsterdam, Netherlands\\
% ...
$ ^{45}$Instituto Galego de F{\'\i}sica de Altas Enerx{\'\i}as (IGFAE), Universidade de Santiago de Compostela, Santiago de Compostela, Spain\\
% ...
$ ^{49}$Physik-Institut, Universit{\"a}t Z{\"u}rich, Z{\"u}rich, Switzerland\\
% ...
$ ^{53}$H.H. Wills Physics Laboratory, University of Bristol, Bristol, United Kingdom\\
% ...
$ ^{59}$Oliver Lodge Laboratory, University of Liverpool, Liverpool, United Kingdom\\
% ...
$ ^{64}$University of Cincinnati, Cincinnati, OH, United States\\
% ...
$ ^{67}$Syracuse University, Syracuse, NY, United States\\
% ...
$ ^{83}$University of Michigan, Ann Arbor, United States, associated to $^{67}$\\
\bigskip
% ...
\medskip
}
\end{flushleft}

\end{document}